\documentclass[draft%
]{article}
\usepackage{amsthm}
\usepackage[]{amsmath}
\usepackage{algorithm}
\usepackage{algpseudocode}

\newtheorem{lem}{Lemma}[]
\newtheorem{thm}[lem]{Theorem}

\renewcommand{\vec}[1]{\ensuremath{\mathbf{#1}}}

\title{Finite automata for testing uniqueness of Eulerian trails}
\author{Qiang Li \\ T-Life Research Center, Fudan University, Shanghai, China
\and Hui-Min Xie \\ Department of Mathematics, Suzhou University, Suzhou, China}
\date{July~17, 2005}

\begin{document}
\maketitle
\begin{abstract}
  We investigate the condition under which the Eulerian trail of a digraph is unique, and design a finite automaton to examine it. The algorithm is effective, for if the condition is violated, it will be noticed immediately without the need to trace through the whole trail.
\end{abstract}

\section{Introduction}
The problem of finding an Eulerian trail in a traversable directed pseudograph is well solved, and a counting formula is given in~\cite{Kandel1996,Hao2001}. But in some applications, like reconstructing a string from its composition of short substrings, as discussed in various contexts~\cite{Pevzner1989,Hao2001,Kontorovich2004,Chauhan2004}, uniqueness rather than the exact number is mostly cared about, so the tedious calculation seems unnecessary. Considering a trail as a symbolic sequence over the set of vertices, Kontorovich showed that the unique Eulerian trails form a regular language~\cite{Kontorovich2004}. We present a different proof by characterizing its complement, which leads to an effective implementation of a deterministic finite automaton (DFA) that accepts it, and gain an insight into its structure from the aspect of minimal forbidden words.

\section{Results}
In the following, we will freely switch the concepts from the theories of graph and formal language, and when the latter viewpoint is emphasized, the set of vertices $V$ is noted $\Sigma$.

\subsection{The language}
Pevzner~\cite{Pevzner2000} proved that any two Eulerian trails of a digraph $G$ can be transformed into each other by a series of operations called \emph{rotations} and \emph{transpositions}. Roughly speaking, rotations correspond to the choice of initial vertex if the trail is closed, and a transposition swaps the order of two paths between a pair of vertices in the trail. Not losing generality, we always suppose that the initial vertex is fixed. Thus an Eulerian trail is not unique only if it has a transposition
\begin{equation}\label{eqn:Tab}
  T: uaxbzaybv\rightarrow uaybzaxbv,
\end{equation}
where $a, b \in \Sigma$, and $u, v, x, y, z \in \Sigma^*$. If $a=b$, it degenerates to the form
\begin{equation}\label{eqn:Taa}
  T:uaxayav\rightarrow uayaxav.
\end{equation}

On the other hand, only $x\neq y$ does not assure that the transposition makes a trail different, e.g.\ let $x=ba$ and $u=v=y=z=\epsilon$, then the trail in \eqref{eqn:Tab} becomes $t=ababab$, which is invariant under the operation, and is actually unique. To eliminate this case, we further request that the two $a$'s before $x$ and $y$ on the left hand side of \eqref{eqn:Tab} or \eqref{eqn:Taa} are followed by distinct vertices. Then we call the corresponding transposition to be \emph{proper}.

\begin{lem}
  Every non-identical transposition is equivalent to a proper transposition.
\end{lem}
\begin{proof}
  For any transposition $T(t)\neq t$, we can write it in the form of \eqref{eqn:Tab} or \eqref{eqn:Taa}. If both $a$'s are followed by $a'$, then let $u'=ua$.
  \begin{enumerate}
    \item If $a\neq b$, then $t=u'xbzaybv$, where $x\neq y$.
      \begin{enumerate}
	\item\label{case:xy} If $x\neq\epsilon$ and $y\neq\epsilon$, then we can write $x=a'x'$ and $y=a'y'$, and let $z'=za$. Otherwise, $a'=b$.
	\item If $x=\epsilon$, then we can write $y=a'y'$, and let $x'=za$.
	\item If $y=\epsilon$, then we can write $x=a'x'$, and let $y'=za$.
      \end{enumerate}
    \item If $a=b$, then $t=u'xayav$, where $x\neq y$.
      \begin{enumerate}
	\item If $x\neq\epsilon$ and $y\neq\epsilon$, then we can write $x=a'x'$ and $y=a'y'$. Otherwise, $a'=a$.
	\item If $x=\epsilon$, then we can write $y=a'y'$, and let $x'=\epsilon$.
	\item If $y=\epsilon$, then we can write $x=a'x'$, and let $y'=\epsilon$.
      \end{enumerate}
  \end{enumerate}
  Therefore, $t$ has a transposition $T':u'a'x'bz'a'y'bv\rightarrow u'a'y'bz'a'x'bv$ in case (\ref{case:xy}) or $T':u'a'x'a'y'a'v\rightarrow u'a'y'a'x'a'v$ in the other cases. Note $T'(s)=T(s)$. Substitute $T'$ for $T$ and repeat the above process, we will eventually get an equivalent proper transposition.
\end{proof}

We conclude that an Eulerian trail $t$ is unique if and only if it does not have a proper transposition. Let $L$ be the language composed of unique Eulerian trails and $L'$ be the language composed of those with proper transpositions, then they are complementary to each other.

By the definition of proper transposition, all sequences in $L'$ have a unified form
\begin{equation}
  t=uawaybv,
  \label{eq:fw}
\end{equation}
where $u,v,w,y\in V^*$, $b$ appears in $aw$, and the vertices next to the two $a$'s are distinct. It results in a right-linear grammar $G$ that generates $L'$:
\begin{align*}
  S&\rightarrow dS|aA_a,\\
  A_a&\rightarrow cB_{acc}|aC_{aa},\\
  B_{acb}&\rightarrow dB_{acb}|dB_{acd}|aC_{cb},\\
  C_{cb}&\rightarrow dD_b\ (d\neq c)|bR\ (b\neq c),\\
  D_b&\rightarrow dD_b|bR,\\
  R&\rightarrow dR|\epsilon,
\end{align*}
where $a, b, c, d$ run over $\Sigma$. Therefore, $L'$ is a regular language, and $L=\overline{L'}$ is also regular.

\subsection{The finite automaton}
Technically we can construct a finite automaton that accepts $L$ from $G(L')$, but it is more convenient to design it directly, like the following.

\paragraph{Input alphabet}
\[\Sigma = V.\]

\paragraph{States}
\[Q = P \times N \times C,\]
where
\begin{itemize}
  \item $P = \Sigma\cup\{a_0\}$, where $a_0\notin\Sigma$ denotes the beginning of the sequence, records the last inputed vertex,
  \item $N = (\Sigma\cup\{\epsilon\})^{m+1}$, where $m=|V|$, records the latest followings of every vertex including $a_0$,
  \item $C = \{\mathtt{WHITE},\mathtt{BLACK}\}^m$ is the ``color'' of every vertex. A vertex is colored black if it is in a circuit $awa$, where the vertex following the tail $a$ differs from that of the head $a$.
\end{itemize}

\paragraph{Initial state}
\[
\vec{q}_0 = (a_0, \epsilon ^{m+1}, \mathtt{WHITE}^m).
\]

\paragraph{Final states}
\[
F = \{(p, \vec{n}, \vec{c}) \in Q\ |\ \vec{c} \neq \mathtt{BLACK}^m\}.
\]

\paragraph{Transition function}
\begin{algorithmic}[1]
  \Procedure{$\delta$}{$\vec{q}, a$}
  \If{$n_p \neq \epsilon$ and $n_p \neq a$}
    \State $b \gets p$
    \Repeat
      \State $c_b \gets \mathtt{BLACK}$
      \State $b \gets n_b$
    \Until{$b = p$}
  \EndIf
  \If{$c_a = \mathtt{BLACK}$}
    \State $\vec{c} \gets \mathtt{BLACK}^m$
  \EndIf
  \State $n_p \gets a$
  \State $p \gets a$
  \EndProcedure
\end{algorithmic}

Now we prove that the DFA $M=(Q,\Sigma,\delta,\vec{q}_0,F)$ accepts $L$.
\begin{proof}
  First we show $L(M)\subset L$ by proving its contrapositive. If $t\notin L$, then it has the form of \eqref{eq:fw}. The design of $M$ assures that $\vec{c}$ becomes $\mathtt{BLACK}^m$ after $b$ is inputed and remains so, thus $M$ does not accept $t$.

  Then we prove $L \subset L(M)$ by induction on the length of the input sequence $t$.

  Basis: For $|t|=0$, $t=\epsilon\in L$. Since $\vec{q}_0\in F$, $t\in L(M)$.

  Induction: For $|t|>0$, if $t=sa\in L$, then $s\in L$, and by the inductive hypothesis $s\in L(M)$. We prove $t\in L(M)$ by contradiction. Assume to the contrary that $t\notin L(M)$, then there are two cases:
  \begin{enumerate}
    \item If $c_a=\mathtt{BLACK}$ just after $s$ is inputed, then $s$ must have the form $ubwby$, where $a$ appears in $bw$ and the vertices following the two $b$'s are distinct. Thus $sa\in L'$, which contradicts $t\in L$.
    \item If $c_a=\mathtt{WHITE}$ just after $s$ is inputed, then $s$ must have the form $upwp$, where $a$ appears in $pw$ and the vertex following the first $p$ is not $a$. Again $sa\in L'$, which contradicts $t\in L$.
  \end{enumerate}
  We conclude that $L(M)=L$.
\end{proof}

\subsection{Minimal forbidden words}
Since $L$ is a factorial language, i.e.\ for any $t\in L$, all factors of $t$ also belong to $L$, it can be determined by its \emph{minimal forbidden words} (MFW)~\cite{Xie1996}. A string $r$ is a minimal forbidden word of $L$ if $r\notin L$ while all the factors of $r$ belong to $L$. We categorize $\mathtt{MFW}(L)$ into sequences in the following two forms, which compose a language $L''$:
\begin{align}
  r&=axbzayb, a\neq b,\label{mfw:ab}\\
  r&=axaya,\label{mfw:aa}
\end{align}
where
\begin{enumerate}
  \item \label{cond:empty}$x\neq\epsilon$ or $y\neq\epsilon$,
  \item \label{cond:l}$x, y, z\in L$,
  \item \label{cond:inter}$x, y, z$ do not contain $a, b$, and each two of $x, y, z$ do not contain common vertices.
\end{enumerate}
\begin{thm}
  $L''=\mathtt{MFW}(L)$.
\end{thm}
\begin{proof}
  By definition all words in $L''$ are minimal forbidden words. Then we prove that $L''$ is complete, i,e.\ $L'\subset\Sigma^*L''\Sigma^*$. For any $t\in L'$, it must has a form of~\eqref{eq:fw}, then $r=awayb\notin L$ satisfies the condition~\ref{cond:empty}. If it violates the condition~\ref{cond:l}, e.g.\ $x\notin L$, then let $t=x$. Repeat the above process until the condition~\ref{cond:l} holds. Then if $y$ contains a vertex $c$ which appears in $aw$, $t$ must have a prefix $away'c\notin L$. Therefore, $t$ has a word $r$ in the form \eqref{mfw:ab} or \eqref{mfw:aa} where $y$ does not contain $a, b$ or common vertex with $x, z$. Since reversing every edge's direction in a graph does not changes the number of its Eulerian trails, $L$ is reversal. So we can also request that $x$ does not contain $a, b$ or common vertex with $z$.
\end{proof}

We can determine $L''$ by recursion on $|\Sigma|$. For the simplest non-trivial case, say $\Sigma=\{0,1\}$, $L''$ can be represented by a regular expression $001^+0+01^+00+110^+1+10^+11$.

\section*{Acknowledgement}
QL would like to thank Prof. Bai-Lin~Hao for asking the probability of an Eulerian trail to be unique in some classes of graphs, and thanks Chan~Zhou for indicating the reference~\cite{Pevzner2000} for him.

\bibliography{unieuler}
\bibliographystyle{plain}
\end{document}